\documentclass{article} \usepackage{ltwol2e} \usepackage{epsfig}
\def\gsim{\ \rlap{\raise 3pt \hbox{$>$}}{\lower 3pt \hbox{$\sim$}}\ }
\def\lsim{\ \rlap{\raise 3pt \hbox{$<$}}{\lower 3pt \hbox{$\sim$}}\ }

\begin{document}

\begin{titlepage}

\begin{flushright}
CERN-TH/98-300\\ 
hep-ph/9809343
\end{flushright}

\vspace{2.5cm}

\begin{center}
\Large\bf  Light Rays of New Physics: CP Violation\\  in
\boldmath$B\to X_s\gamma$\unboldmath\ Decays
\end{center}

\vspace{1.2cm}

\begin{center}
Matthias Neubert\\ {\sl Theory Division, CERN, CH-1211 Geneva 23,
Switzerland}
\end{center}

\vspace{1.3cm}

\begin{center}
{\bf Abstract:}\\[0.3cm]
\parbox{11cm}{ 
The observation of a sizable direct CP asymmetry in the
inclusive decays $B\to X_s\gamma$ would be a clean signal of New
Physics. In the Standard Model, this asymmetry is below 1\% in
magnitude. In extensions of the Standard Model with new CP-violating
couplings, large asymmetries are possible without conflicting with the
experimental value of the $B\to X_s\gamma$ branching ratio. In
particular, large asymmetries arise naturally in models with enhanced
chromo-magnetic dipole transitions. Some generic examples of such
models are explored and their implications for the semileptonic
branching ratio and charm yield in $B$ decays discussed.}
\end{center}

\vspace{1cm}

\begin{center}
{\sl To appear in the Proceedings of the QCD 98 Euroconference\\
Montpellier, France, 2--8 July 1998\\ and\\ XXIXth International
Conference on High Energy Physics\\ Vancouver, B.C., Canada, 23--29
July 1998}
\end{center}

\vfil
\noindent
CERN-TH/98-300\\ 
September 1998

\end{titlepage}

\setcounter{page}{1}

\title{LIGHT RAYS OF NEW PHYSICS: CP VIOLATION IN  \boldmath$B\to
X_s\gamma$\unboldmath\ DECAYS}

\author{M. NEUBERT}

\address{Theory Division, CERN, CH-1211 Geneva 23, Switzerland\\
E-mail: Matthias.Neubert@cern.ch}

\twocolumn[\maketitle\abstracts{ The observation of a sizable direct
CP asymmetry in the inclusive decays $B\to X_s\gamma$ would be a clean
signal of New Physics. In the Standard Model, this asymmetry is below
1\% in magnitude. In extensions of the Standard Model with new
CP-violating couplings, large asymmetries are possible without
conflicting with the experimental value of the $B\to X_s\gamma$
branching ratio. In particular, large asymmetries arise naturally in
models with enhanced chromo-magnetic dipole transitions. Some generic
examples of such models are explored and their implications for the
semileptonic branching ratio and charm yield in $B$ decays discussed.}]

\section{Introduction}

Studies of rare decays of $B$ mesons have the potential to uncover the
origin of CP violation and provide hints to physics beyond the
Standard Model. The measurements of several CP asymmetries will make
it possible to test whether the CKM paradigm is correct, or whether
additional sources of CP violation are required. In order to achieve
this goal, it is necessary that the theoretical calculations of
CP-violating observables are, to a large extent, free of hadronic
uncertainties. This can be achieved, e.g., by measuring time-dependent
asymmetries in the decays of neutral $B$ mesons into particular CP
eigenstates. In many other cases, however, the theoretical predictions
for direct CP violation in exclusive decays are obscured by
strong-interaction effects \cite{Blok96}$^-$\cite{At97}.

Inclusive decay rates of $B$ mesons, on the other hand, can be
reliably calculated in QCD using the operator product expansion. Up to
small bound-state corrections these rates agree with the parton model
predictions for the underlying decays of the $b$ quark. The
disadvantage that the sum over many final states partially dilutes the
CP asymmetries in inclusive decays is compensated by the fact that,
because of the short-distance nature of these processes, the strong
phases are calculable using quark--hadron duality. In this talk, I
report on a study \cite{Alex1} of direct CP violation in the rare
radiative decays $B\to X_s\gamma$, both in the Standard Model and
beyond. These decays have already been observed experimentally, and
copious data samples will be collected at the $B$ factories. The
theoretical analysis relies only on the weak assumption of global
quark--hadron duality, and the leading nonperturbative corrections are
well understood.

We perform a model-independent analysis of CP-violating effects  in
terms of the effective Wilson coefficients $C_7\equiv C_7^{\rm
eff}(m_b)$ and $C_8\equiv C_8^{\rm eff}(m_b)$ multiplying the
(chromo-) magnetic dipole operators $O_7=e\,m_b\,\bar
s_L\sigma_{\mu\nu} F^{\mu\nu} b_R$ and $O_8=g_s m_b\,\bar
s_L\sigma_{\mu\nu} G^{\mu\nu} b_R$ in the effective weak Hamiltonian,
allowing for generic New Physics contributions to these coefficients.
Several extensions of the Standard Model in which such contributions
arise have been explored, e.g., in Refs.~7--11.
We find that in the Standard Model the direct CP asymmetry in $B\to
X_s\gamma$ decays is very small (below 1\% in magnitude) because of a
combination of CKM and GIM suppression, both of which can be lifted in
New Physics scenarios with additional contributions to the dipole
operators containing new weak phases. We thus propose a measurement of
the inclusive CP asymmetry in radiative $B$ decays as a clean and
sensitive probe of New Physics. Studies of direct CP violation in the
inclusive decays $B\to X_s\gamma$ have been performed previously by
several authors, both in the Standard Model \cite{Soares} and in
certain extensions of it \cite{Wolf,Asat}. In all cases rather small
asymmetries were obtained. We generalize and extend these analyses in
various ways. Besides including some contributions neglected in
previous works, we investigate a class of New Physics models with
enhanced chromo-magnetic dipole contributions, in which large CP
asymmetries of order 10--50\% are possible. We also employ a
next-to-leading order analysis of the CP-averaged $B\to X_s\gamma$
branching ratio in order to derive constraints on the parameter space
of the New Physics models considered.

\boldmath
\section{Direct CP violation in $B\to X_s\gamma$ decays}
\unboldmath
\label{sec:ACP}

The starting point in the calculation of the inclusive $B\to
X_s\gamma$ decay rate is provided by the effective weak Hamiltonian
renormalized at the scale $\mu=m_b$. Direct CP violation in these
decays may arise from the interference of non-trivial weak phases,
contained in CKM parameters or in possible New Physics contributions
to the Wilson coefficient functions, with strong phases provided by
the imaginary parts of the matrix elements of the operators in the
effective Hamiltonian \cite{Band}. These imaginary parts first arise
at $O(\alpha_s)$ from loop diagrams containing charm quarks, light
quarks or gluons. Using the formulae of Greub et al.\ for these
contributions~\cite{Greub}, we calculate at next-to-leading order the
difference $\Delta\Gamma=\Gamma(\bar B\to X_s\gamma)-\Gamma(B\to
X_{\bar s}\gamma)$ of the CP-conjugate, inclusive decay rates. The
contributions to $\Delta\Gamma$ from virtual corrections arise from
interference of the one-loop diagrams with insertions of the operators
$O_2$ and $O_8$ with the tree-level diagram containing $O_7$. Here
$O_2=\bar s_L\gamma_\mu q_L\,\bar q_L\gamma^\mu b_L$ with $q=c,u$ are
the usual current--current operators in the effective Hamiltonian.
There are also contributions to $\Delta\Gamma$ from gluon
bremsstrahlung diagrams with a charm-quark loop. They can interfere
with the tree-level diagrams for $b\to s\gamma g$ containing an
insertion of $O_7$ or $O_8$. Contrary to the virtual corrections, for
which in the parton model the photon energy is fixed to its maximum
value, the gluon bremsstrahlung diagrams lead to a non-trivial photon
spectrum, and so the results depend on the experimental lower cutoff
on the photon energy. We define a quantity $\delta$ by the requirement
that $E_\gamma > (1-\delta) E_\gamma^{\rm max}$. Combining the two
contributions and dividing the result by the leading-order expression
for twice the CP-averaged inclusive decay rate, we find for the CP
asymmetry
\begin{eqnarray}
   A_{\rm CP}^{b\to s\gamma}(\delta) &=& \frac{\Gamma(\bar B\to
   X_s\gamma)-\Gamma(B\to X_{\bar s}\gamma)} {\Gamma(\bar B\to
   X_s\gamma)+\Gamma(B\to X_{\bar s}\gamma)}
   \Bigg|_{E_\gamma>(1-\delta) E_\gamma^{\rm max}} \nonumber\\ &=&
   \frac{\alpha_s(m_b)}{|C_7|^2}\,\Bigg\{ \frac{40}{81}\,\mbox{Im}[C_2
   C_7^*] - \frac 49\,\mbox{Im}[C_8 C_7^*] \nonumber\\ &&\mbox{}-
   \frac{8z}{9}\,\Big[ v(z) + b(z,\delta) \Big]\,
   \mbox{Im}[(1+\epsilon_s) C_2 C_7^*] \nonumber\\ &&\mbox{}+
   \frac{8z}{27}\,b(z,\delta)\, \mbox{Im}[(1+\epsilon_s) C_2 C_8^*]
   \Bigg\} \,,
\label{ACP}
\end{eqnarray}
where $z=(m_c/m_b)^2$, and the explicit expressions for the functions
$g(z)$ and $b(z,\delta)$ can be found in Ref.~6.
The quantity $\epsilon_s$ is a ratio of CKM matrix elements given by
\begin{equation}
   \epsilon_s = \frac{V_{us}^* V_{ub}}{V_{ts}^* V_{tb}} \approx
   \lambda^2 (i\eta-\rho) = O(10^{-2}) \,,
\end{equation}
where $\lambda=\sin\theta_{\rm C}\approx 0.22$ and $\rho,\eta=O(1)$
are the Wolfenstein parameters. An estimate of the $C_2$--$C_7$
interference term in (\ref{ACP}) was obtained previously by Soares
\cite{Soares}, who neglected the contribution of $b(z,\delta)$ and
used an approximation for the function $v(z)$. The relevance of the
$C_8$--$C_7$ interference term for two-Higgs-doublet models, and for
left--right symmetric extensions of the Standard Model, was explored
in Refs.~13,14.

In the Standard Model, the Wilson coefficients take the real values
$C_2\approx 1.11$, $C_7\approx -0.31$ and $C_8\approx -0.15$. The
imaginary part of the small quantity $\epsilon_s$ is thus the only
source of CP violation. All terms involving this quantity are GIM
suppressed by a power of the small ratio $z=(m_c/m_b)^2$, reflecting
the fact that there is no non-trivial weak phase difference in the
limit where $m_c=m_u=0$. Hence, the Standard Model prediction for the
CP asymmetry is suppressed by three small factors: $\alpha_s(m_b)$
arising from the strong phases, $\sin^2\!\theta_{\rm C}$ reflecting
the CKM suppression, and $(m_c/m_b)^2$ resulting from the GIM
suppression. The numerical result for the asymmetry depends on the
values of the strong coupling constant and the ratio of the
heavy-quark pole masses, for which we take $\alpha_s(m_b)\approx
0.214$ (corresponding to $\alpha_s(m_Z)=0.118$ and two-loop evolution
down to the scale $m_b=4.8$\,GeV) and $\sqrt z=m_c/m_b=0.29$. This
yields $A_{\rm CP,SM}^{b\to s\gamma} \approx
(1.5\mbox{--}1.6)\%\,\eta$ depending on the value of $\delta$. With
$\eta\approx 0.2$--0.4 as suggested by phenomenological analyses, we
find a tiny asymmetry of about 0.5\%, in agreement with the estimate
obtained in Ref.~12.
Expression (\ref{ACP}) applies also to the decays $B\to X_d\,\gamma$,
the only difference being that in this case the quantity $\epsilon_s$
must be replaced with the corresponding quantity $\epsilon_d =
(V_{ud}^* V_{ub})/(V_{td}^* V_{tb}) \approx
(\rho-i\eta)/(1-\rho+i\eta) = O(1)$. Therefore, in the Standard Model
the CP asymmetry in $B\to X_d\,\gamma$ decays is larger by a factor of
about $-20$ than that in $B\to X_s\gamma$ decays. However,
experimentally it is difficult to distinguish between $B\to X_s\gamma$
and $B\to X_d\,\gamma$ decays. If only their sum is measured, the CP
asymmetry vanishes by CKM unitarity \cite{Soares}.

\begin{table}
\caption{Values of the coefficients $a_{ij}$ in \%}
\vspace{0.2cm}
\begin{center}
\begin{tabular}{|cc|ccc|}
\hline $\delta$ & $E_\gamma^{\rm min}~[{\rm GeV}]$ & $a_{27}$ &
$a_{87}$ & $a_{28}$ \\ \hline\hline 1.00 & 0.00 & 1.06 & $-9.52$ &
0.16 \\ 0.30 & 1.85 & 1.23 & $-9.52$ & 0.10 \\ 0.15 & 2.24 & 1.40 &
$-9.52$ & 0.04 \\ \hline
\end{tabular}
\end{center}
\label{tab:aij}
\end{table}

{}From (\ref{ACP}) it is apparent that two of the suppression factors
operative in the Standard Model, $z$ and $\lambda^2$, can be avoided
in models where the effective Wilson coefficients $C_7$ and $C_8$
receive additional contributions involving non-trivial weak
phases. Much larger CP asymmetries then become possible. In order to
investigate such models, we may to good approximation neglect the
small quantity $\epsilon_s$ and write
\begin{eqnarray}
   A_{\rm CP}^{b\to s\gamma}(\delta)  &=&
   a_{27}(\delta)\,\mbox{Im}\!\left[ \frac{C_2}{C_7} \right] +
   a_{87}\,\mbox{Im}\!\left[ \frac{C_8}{C_7} \right] \nonumber\\
   &&\mbox{}+ a_{28}(\delta)\,\frac{\mbox{Im}[C_2 C_8^*]}{|C_7|^2} \,.
\label{3terms}
\end{eqnarray}
The values of the coefficients $a_{ij}$ are shown in
Table~\ref{tab:aij} for three choices of the cutoff on the photon
energy: $\delta=1$ corresponding to the (unrealistic) case of a fully
inclusive measurement, $\delta=0.3$ corresponding to a restriction to
the part of the spectrum above $1.85$\,GeV, and $\delta=0.15$
corresponding to a cutoff that removes almost all of the background
from $B$ decays into charmed hadrons. In practice, a restriction to
the high-energy part of the photon spectrum is required for
experimental reasons. Note, however, that the result for the CP
asymmetry is not very sensitive to the choice of the cutoff. Whereas
the third term in (\ref{3terms}) is generally very small, the first
two terms can give rise to sizable effects. Assume, e.g., that there
is a New Physics contribution to $C_7$ of similar magnitude as the
Standard Model contribution (so as not to spoil the prediction for the
$B\to X_s\gamma$ branching ratio) but with a non-trivial weak
phase. Then the first term in (\ref{3terms}) may give a contribution
of up to about 5\% in magnitude. Similarly, if there are New Physics
contributions to $C_7$ and $C_8$ such that the ratio $C_8/C_7$ has a
non-trivial weak phase, the second term may give a contribution of up
to about $10\%\times|C_8/C_7|$. In models with a strong enhancement of
$|C_8|$ with respect to its Standard Model value, there is thus the
possibility of generating a large direct CP asymmetry in $B\to
X_s\gamma$ decays.

The impact of nonperturbative power corrections on the rate ratio
defining the CP asymmetry is very small, since most of the corrections
cancel between the numerator and the denominator.  Potentially the
most important bound-state effect is the Fermi motion of the $b$ quark
inside the $B$ meson, which determines the shape of the photon energy
spectrum in the endpoint region. This effect is included in the
heavy-quark expansion by resumming an infinite set of leading-twist
contributions into a ``shape function'', which governs the momentum
distribution of the heavy quark inside the meson \cite{shape,Dike95}.
The physical decay distributions are obtained from a convolution of
parton model spectra with this function. In the process, phase-space
boundaries defined by parton kinematics are transformed into the
proper physical boundaries defined by hadron kinematics. Details of
the implementation of this effect can be found in Refs.~6,19.
We note  that the largest coefficient, $a_{87}$, is not affected by
Fermi  motion, and the impact on the other two coefficients  is rather
mild.  As a consequence, the predictions for the CP asymmetry are
quite  insensitive to bound-state effects, even if a restriction on
the  high-energy part of the photon spectrum is imposed.

Below we explore the structure of New Physics models with a
potentially large inclusive CP asymmetry. A non-trivial constraint on
such models is that they must yield an acceptable result for the
total, CP-averaged $B\to X_s\gamma$ branching ratio, which has been
measured experimentally. Taking a weighed average of the results
reported by the CLEO and ALEPH Collaborations \cite{CLEO,ALEPH} at
this Conference gives  $\mbox{B}(B\to X_s\gamma)=(3.14\pm 0.48)\times
10^{-4}$. The complete  theoretical prediction for the $B\to
X_s\gamma$ branching ratio at next-to-leading order has been presented
for the first time by Chetyrkin et al.\ \cite{Chet}, and subsequently
has been discussed by several authors
\cite{Buras}$^-$\cite{newGreub}. It depends on the Wilson coefficients
$C_2$, $C_7$ and $C_8$ through the combinations $\mbox{Re}[C_i
C_j^*]$.  Recently, we have extended these analyses in several
aspects, including a discussion of Fermi motion effects and a
conservative analysis of perturbative uncertainties \cite{newpaper}.
In contrast to the case of the CP asymmetry, Fermi motion effects are
very important when comparing experimental data for the $B\to
X_s\gamma$ branching ratio with theoretical predictions. With our
choice of parameters, we obtain for the total branching ratio in the
Standard Model $\mbox{B}(B\to X_s\gamma)=(3.29\pm 0.33)\times
10^{-4}$, which is in very good agreement with the experimental
findings.

\section{CP asymmetry beyond the Standard Model}

In order to explore the implications of various New Physics scenarios
for the CP asymmetry and branching ratio in $B\to X_s\gamma$ decays,
we use the renormalization group to express the Wilson coefficients
$C_7=C_7^{\rm eff}(m_b)$ and $C_8=C_8^{\rm eff}(m_b)$ in terms of
their  values at the high scale $m_W$, for which we write
$C_{7,8}(m_W)=  C_{7,8}^{\rm SM}(m_W)+C_{7,8}^{\rm new}(m_W)$. The
first term  corresponds to the Standard Model contributions, which are
functions of  the mass ratio $x_t=(m_t/m_W)^2$. Numerically, one
obtains
\begin{eqnarray}
   C_7 &\approx& -0.31 + 0.67\,C_7^{\rm new}(m_W) + 0.09\,C_8^{\rm
    new}(m_W) \,, \nonumber\\ C_8 &\approx& -0.15 + 0.70\,C_8^{\rm
    new}(m_W) \,.
\label{C7C8}
\end{eqnarray}
We choose to parametrize our results in terms of the magnitude and
phase of the New Physics contribution $C_8^{\rm new}(m_W)\equiv
K_8\,e^{i\gamma_8}$ as well as the ratio
\begin{equation}
   \xi = \frac{C_7^{\rm new}(m_W)}{Q_d\,C_8^{\rm new}(m_W)} \,,
\label{xidef}
\end{equation}
where $Q_d=-\frac 13$. A given New Physics scenario predicts these
quantities at some large scale $M$. Using the renormalization group,
it is then possible to evolve these predictions down to the scale
$m_W$.  Typically, $\xi\equiv\xi(m_W)$ tends to be smaller than
$\xi(M)$ by an amount of order $-0.1$ to $-0.3$, depending on how
close the New Physics is to the electroweak scale \cite{Alex1}. We
restrict ourselves to cases where the parameter $\xi$ in (\ref{xidef})
is real; otherwise there would be even more potential for CP
violation. This happens if there is a single dominant New Physics
contribution, such as the virtual exchange of a new heavy particle,
contributing to both the magnetic and the chromo-magnetic dipole
operators.

\begin{table}
\caption{Ranges of $\xi(M)$ for various New Physics contributions to
$C_7$ and $C_8$, characterized by the particles in penguin diagrams}
\vspace{0.2cm}
\begin{center}
\begin{tabular}{|lc|}
\hline Class-1 models & $\xi(M)$ \\ \hline\hline neutral
scalar--vectorlike quark & 1 \\ gluino--squark ($m_{\tilde g} < 1.37
m_{\tilde q}$) & $-(0.13\mbox{--}1)$ \\ techniscalar & $\approx-0.5$
\\ \hline\hline Class-2 models & $\xi(M)$ \\ \hline\hline scalar
diquark--top & 4.8--8.3 \\ gluino--squark ($m_{\tilde g} > 1.37
m_{\tilde q}$) & $-(1\mbox{--}2.9)$ \\ charged Higgs--top &
$-(2.4\mbox{--}3.8)$ \\ left--right $W$--top & $\approx -6.7$ \\
Higgsino--stop & $-(2.6\mbox{--}24)$ \\ \hline
\end{tabular}
\end{center}
\label{tab:xi}
\end{table}

\begin{figure}
\epsfxsize=8.7cm \centerline{\epsffile{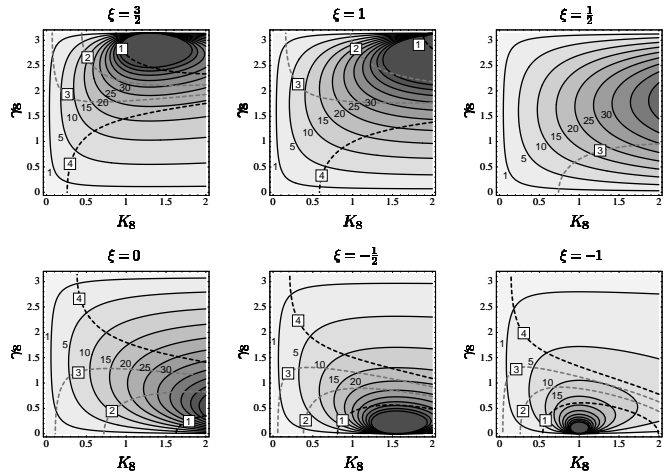}}
\caption{Contours for the CP asymmetry $A_{\rm CP}^{b\to s\gamma}$ for
various class-1 models. We show contours only until values  $A_{\rm
CP}=50\%$; for such large values, the theoretical expression
(\protect\ref{3terms}) for the asymmetry would have to be extended to
higher orders to get a reliable result.}
\label{fig:models1}
\end{figure}

Ranges of $\xi(M)$ for several illustrative New Physics scenarios are
collected in Table~\ref{tab:xi}. For a detailed discussion of the
model parameters which lead to the $\xi$ values quoted in the table
the reader is referred to Ref.~6.
Our aim is not to carry out a detailed study of each model, but to
give an idea of the sizable variation that is possible in $\xi$. It is
instructive to distinguish two classes of models: those with moderate
(class-1) and those with large (class-2) values of $|\xi|$. It follows
from (\ref{C7C8}) that for small positive values of $\xi$ it is
possible to have large complex contributions to $C_8$ without
affecting too much the magnitude and phase of $C_7$, since
\begin{equation}
   \frac{C_8}{C_7}\approx\frac{0.70 K_8\,e^{i\gamma_8}-0.15}
    {(0.09-0.22\xi) K_8\,e^{i\gamma_8}-0.31} \,.
\label{C7C8rat}
\end{equation}
This is also true for small negative values of $\xi$, albeit over a
smaller region of parameter space. New Physics scenarios that have
this property belong to class-1 and have been explored in Ref.~8.
They allow for large CP asymmetries resulting from the $C_7$--$C_8$
interference term in (\ref{3terms}).  Figure~\ref{fig:models1} shows
contour plots for the CP asymmetry in the $(K_8,\gamma_8)$ plane for
six different choices of $\xi$ between $\frac32$ and $-1$, assuming a
cutoff $E_\gamma>1.85$\,GeV on the photon energy. For each value of
$\xi$, the plots cover the region $0\le K_8\le 2$ and
$0\le\gamma_8\le\pi$ (changing the sign of $\gamma_8$ would only
change the sign of the CP asymmetry). The contour lines refer to
values of the asymmetry of 1\%, 5\%, 10\%, 15\% etc. The dashed lines
indicate contours where the $B\to X_s\gamma$ branching ratio takes
values between $1\times 10^{-4}$ and $4\times 10^{-4}$, as indicated
by the numbers inside the squares.  The Standard Model prediction with
this choice of the photon-energy cutoff is about $3\times
10^{-4}$. The main conclusion to be drawn from the figure is that in
class-1 scenarios there is a great potential  for having a sizable CP
asymmetry in a large region of parameter space.  Any point to the
right of the 1\% contour for $A_{\rm  CP}^{b\to s\gamma}$ cannot be
accommodated by the Standard Model. Note that quite generally the
regions of parameter space that yield large values for the CP
asymmetry are not excluded by the experimental constraint on the
CP-averaged branching ratio. To have a large CP asymmetry the products
$C_i C_j^*$ are required to have large imaginary parts, whereas the
total branching ratio is sensitive to the real parts of these
quantities.

There are also scenarios in which the parameter $\xi$ takes on larger
negative or positive values. In such cases, it is not possible to
increase the magnitude of $C_8$ much over its Standard Model value,
and the only way to get a large CP asymmetry from the $C_7$--$C_8$ or
$C_7$--$C_2$ interference terms in (\ref{3terms}) is to have $C_7$
tuned to be very small; however, this possibility is constrained by
the fact that the total $B\to X_s\gamma$ branching ratio must be of an
acceptable magnitude. That this condition starts to become a limiting
factor is already seen in the plots corresponding to $\xi=-\frac12$
and $-1$ in Figure~\ref{fig:models1}. For even larger values of
$|\xi|$, the $C_7$--$C_8$ interference term becomes ineffective,
because the weak phase tends to cancel in the ratio $C_8/C_7$. Then
the $C_2$--$C_7$ interference term becomes the main source of CP
violation; however, as discussed in Section~\ref{sec:ACP}, it cannot
lead to an asymmetry exceeding the level of about 5\% without
violating the constraint that the $B\to X_s\gamma$ branching ratio not
be too small.  Models of this type belong to the class-2 category. The
branching-ratio  constraint allows larger values of $C_8$ for positive
$\xi$. For example, for $\xi\approx 5$, which can be obtained from
scalar diquark--top penguins, asymmetries of 5--20\% are still
consistent with the $B\to X_s\gamma$ bound. On the other hand, for
$\xi\approx-(2.5\mbox{--}5)$, which includes the multi-Higgs-doublet
models, asymmetries of only a few percent are attainable, in agreement
with the findings of previous authors \cite{Wolf,Asat,newGreub}.

\begin{figure}
\epsfxsize=7.5cm \centerline{\epsffile{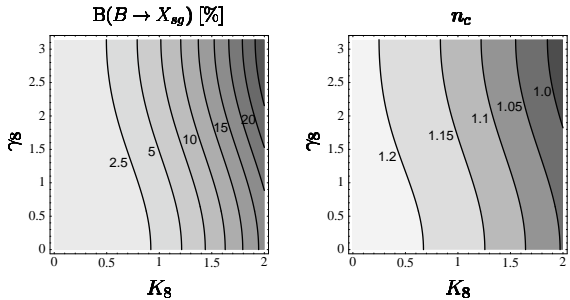}}
\caption{Contours for the $B\to X_{sg}$ branching ratio (left) and for
the charm yield $n_c$ in $B$ decays (right)}
\label{fig:nc}
\end{figure}

The class-1 scenarios explored in Figure~\ref{fig:models1} have the
attractive feature of a possible large enhancement of the magnitude of
the Wilson coefficient $C_8$. This would have important implications
for the phenomenology of the semileptonic branching ratio and charm
yield in $B$ decays, through enhanced production of charmless hadronic
final states induced by the $b\to s g$
transition~\cite{hou}$^-$\cite{CGG95}. At $O(\alpha_s)$, the  $B\to
X_{sg}$ branching ratio is proportional to $|C_8|^2$. The left-hand
plot in Figure~\ref{fig:nc} shows contours for this branching ratio in
the $(K_8,\gamma_8)$ plane. In the Standard Model, $\mbox{B}(B\to
X_{sg})\approx 0.2\%$ is very small; however, in scenarios with
$|C_8|=O(1)$ sizable values of order 10\% for this branching ratio are
possible, which simultaneously lowers the theoretical predictions for
the semileptonic branching ratio and the charm production rate $n_c$
by a factor of $[1+\mbox{B}(B\to X_{sg})]^{-1}$. The current value of
$n_c$ reported by the CLEO Collaboration is \cite{Drell} $1.12\pm
0.05$. Although the systematic errors in this measurement are large,
the result favours values of $\mbox{B}(B\to X_{sg})$ of order
10\%. This is apparent from the right-hand plot in
Figure~\ref{fig:nc}, which shows the central theoretical prediction
for $n_c$ as a function of $K_8$ and $\gamma_8$.  Note that there is
an overall theoretical uncertainty in the value of $n_c$ of about 6\%
resulting from the dependence on quark masses and the renormalization
scale \cite{NeSa}. The theoretical prediction for the semileptonic
branching ratio would have the same dependence on $K_8$ and
$\gamma_8$, with the normalization ${\rm B_{SL}}=(12\pm 1)\%$ fixed at
$K_8=0$. A large value of $\mbox{B}(B\to X_{sg})$ could also help in
understanding the $\eta'$ yields in charmless $B$ decays
\cite{Houeta,Petrov}. For completeness, we note that the CLEO
Collaboration has presented a preliminary upper limit on
$\mbox{B}(B\to X_{sg})$ of 6.8\% (90\% CL)~\cite{Thorn}. It is
therefore worth noting that large CP asymmetries of order 10--20\% can
also be attained at smaller $B\to X_{sg}$ branching ratios of a few
percent, which would nevertheless represent a marked departure from
the Standard Model.

\section{Conclusions}

I have reported on a study of direct CP violation in the inclusive,
radiative decays $B\to X_s\gamma$. From a theoretical point of view,
inclusive decay rates entail the advantage of being calculable in QCD,
so that a reliable prediction for the CP asymmetry can be confronted
with data. From a practical point of view, it is encouraging that
$B\to X_s\gamma$ decays have already been observed experimentally, and
high-statistics measurements will be possible in the near future. We
find that in the Standard Model the CP asymmetry in $B\to X_s\gamma$
decays is strongly suppressed by three small parameters:
$\alpha_s(m_b)$ arising from the necessity of having strong phases,
$\sin^2\!\theta_{\rm C}\approx 5\%$ reflecting a CKM suppression, and
$(m_c/m_b)^2\approx 8\%$ resulting from a GIM suppression. As a
result, the asymmetry is only of order 1\% in magnitude -- a
conclusion that cannot be significantly modified by long-distance
contributions. The  latter two suppression factors are inoperative in
extensions of the  Standard Model for which the effective Wilson
coefficients $C_7$ and  $C_8$ receive additional contributions
involving non-trivial weak  phases. Much larger CP asymmetries are
therefore possible in such  cases.

A model-independent analysis of New Physics scenarios in terms of the
magnitudes and phases of the Wilson coefficients $C_7$ and $C_8$ shows
that sizable CP asymmetries are predicted in large regions of
parameter space. Asymmetries of 10--50\% are possible in models which
allow for a strong enhancement of the coefficient of the
chromo-magnetic dipole operator. They are, in fact, natural unless
there is a symmetry that forbids new weak phases from entering the
Wilson coefficients. Quite generally, having a large CP asymmetry is
not in conflict with the observed value for the CP-averaged $B\to
X_s\gamma$ branching ratio. Indeed, it may help to lower the
theoretical predictions for the semileptonic branching ratio and charm
multiplicity in $B$ decays, thereby bringing these observables  closer
to their experimental values.

The fact that a large inclusive CP asymmetry in $B\to X_s\gamma$
decays is possible in many generic extensions of the Standard Model,
and in a large region of parameter space, offers the possibility of
looking for a signature of New Physics in these decays using data sets
that will become available during the first period of operation of the
$B$ factories. A negative result of such a study would impose
constraints on many New Physics scenarios. A positive signal, on the
other hand, would provide interesting clues about the nature of
physics beyond the Standard Model. In particular, a CP asymmetry
exceeding the level of 10\% would be a strong hint towards enhanced
chromo-magnetic dipole transitions.

We have restricted our analysis to the case of inclusive radiative
decays since they entail the advantage of being theoretically very
clean. However, if there is New Physics that induces a large inclusive
CP asymmetry in $B\to X_s\gamma$ decays it will inevitably also lead
to sizable asymmetries in some related processes. In particular, since
we found that the inclusive CP asymmetry remains almost unaffected if
a cut on the high-energy part of the photon spectrum is imposed, we
expect that a large asymmetry will persist in the exclusive decay mode
$B\to K^*\gamma$, even though a reliable theoretical analysis would be
much more difficult because of the necessity of calculating
final-state rescattering phases \cite{GSW95}. Still, it would be
worthwhile searching for CP violation in this channel.

\newpage
\section*{Acknowledgments}

The work reported here has been done in a most pleasant collaboration
with Alex Kagan, which is gratefully acknowledged.

\end{document}